\documentclass[12ptshowpacs]{revtex4}
\usepackage{amsmath}
\usepackage{amssymb}
\usepackage{graphicx}

\begin{document}

\title{Discrete solitons in an array of quantum dots}
\author{Goran Gligori\'{c}, Aleksandra Maluckov, and Ljup\v{c}o Had\v{z}%
ievski}
\affiliation{Vin\v{c}a Institute of Nuclear Sciences, University of Belgrade, P. O. B.
522,11001 Belgrade, Serbia}
\author{ Gregory Ya. Slepyan and Boris A. Malomed}
\affiliation{Department of Physical Electronics, School of Electrical Engineering,
Faculty of Engineering, Tel Aviv University, Tel Aviv 69978, Israel}

\begin{abstract}
We develop a theory for the interaction of classical light fields with an a
chain of coupled quantum dots (QDs), in the strong-coupling regime, taking
into account the local-field effects. The QD chain is modeled by a
one-dimensional (1D) periodic array of two-level quantum particles with
tunnel coupling between adjacent ones. The local-field effect is taken into
regard as QD depolarization in the Hartree-Fock-Bogoliubov approximation.
The dynamics of the chain is described by a system of two discrete nonlinear
Schr\"{o}dinger (DNLS) equations for local amplitudes of the probabilities
of the ground and first excited states. The two equations are coupled by a
cross-phase-modulation cubic terms, produced by the local-field action, and
by linear terms too. In comparison with previously studied DNLS systems, an
essentially new feature is a phase shift between the intersite-hopping
constants in the two equations. By means of numerical solutions, we
demonstrate that, in this QD chain, Rabi oscillations (RO) self-trap into
stable bright\textit{\ Rabi solitons} or \textit{Rabi breathers}. Mobility
of the solitons is considered too. The related behavior of observable
quantities, such as energy, inversion, and electric-current density, is
given a physical interpretation. The results apply to a realistic region of
physical parameters.
\end{abstract}

\pacs{}
\maketitle

{05.45.Yv; 68.65.Hb; 78.67.Ch; 73.21.-b}

\section{Introduction}

Rabi oscillations (RO) represent oscillating transitions of a
two-level quantum system between its stationary states under the
action of a synchronously oscillating field. The RO effect appears
in the regime of the strong coupling of a quantum oscillator with
the driving field, hence it cannot be considered as a small
perturbation. The effect was for the first time theoretically
predicted by Rabi [1] and experimentally observed by Torrey [2] as a
periodic change (``flip-flops") of the orientation of a nuclear spin
in the magnetic field, in the radiofrequency range. Thereafter, RO
were found in many other physical settings, such as
electromagnetically driven atoms [3], semiconductor quantum dots
(QDs) [4], semiconductor charge qubits [5], spin qubits [6],
superconductive charge qubits based on Josephson junctions [7], etc.
In particular, the interest to RO in different types of
nanostructures is stipulated by potential applications to the design
of elements of quantum logic and quantum memory.

\qquad The significant role of local-field effects (electron-hole
dipole-dipole interactions) in building coherent optical response of the
QDs, even in the weak-coupling regime, was predicted in Refs. [8-11]. In
particular, local fields induce a fine structure of the QD
absorption-emission spectrum [10,11]. The local-field effects enhanced by
the strong light-QD coupling manifest themselves in a number of observable
modifications of the conventional RO picture. Some of them, such as a
bifurcation and essentially anharmonic RO regimes, were predicted in Refs.
[12,13] and thereafter experimentally observed in Ref. [14] in quantum-well
islands. The role of local fields in the formation of excitonic RO in
self-assembling QDs was experimentally studied in Ref. [15], using the
two-pulse photon-echo spectroscopy. As a result, it has been concluded that
local-field effects considerably contribute to the QD optical response in
agreement with the theoretical estimations [11-13].

\qquad The analysis of the local-field effects should be based, in
principle, on the general principles of the quantum many-body theory
[16,17]. However, because a direct solution of this problem is
practically impossible, different phenomenological interpretations
of the local field in the QD and, respectively, different
approximate ways for its theoretical description have been
elaborated. One model (scheme A, in terms of Ref. [18]) exploits the
electrodynamic picture: a depolarization field is formed, making a
difference between the local field inside the QD and the external
acting field, due to the screening of the external field by the
charges induced at QD boundaries. In this model, the total
electromagnetic field is not fully transverse, including a
longitudinal component. In the alternative model (scheme B in terms
of Ref. [18]), only the transverse component is associated with the
electromagnetic field, while the longitudinal part is related to the
electron-hole interactions. Both approaches are equivalent and lead
to identical results. In any case, the local fields should be
introduced self-consistently, parallel to the consideration of the
carrier motion. In the framework of single-particle models, this can
be done by means of the Hartree-Fock-Bogoliubov approximation [19].
As a result, specific nonlinear terms appear in the equations of
electron-hole oscillations in the QD, invalidating the superposition
principle.

In spatially extended systems, such as one-dimensional QD chains, collective
effects come into play. This leads to the spatial propagation of the RO in
the form of plane waves and wave packets, as predicted in Refs. [20-22]. The
spatial walk of the quantum transitions is accompanied by the transfer of
energy, quasimomentum, and electron-electron and electron-photon
correlations. To simplify the description of this quantum transport, the
discrete QD chain was modeled as an effective continuous medium [20-22]. Two
different cases may be distinguished with respect to the quantum structure
of the incident field. In the first, semiclassical, case, the spatial
structure of the classical external field follows the quantum motion of the
particles, but light-matter interactions do not manifest themselves in the
structure of the photonic field. Thus, the electron-photon wave function is
factorizable, and only its electronic part should be explicitly examined. As
a result, the RO propagation is described by a linear system of two coupled
Schr\"{o}dinger equations for complex envelope amplitudes of the ground- and
excited-state probabilities [20]. The second case is a truly quantum one. In
that case, the electron-photon exchange associated with the RO propagation
process is accompanied by the transformation of the field distribution over $%
\left\vert n\right\rangle $-photon states, leading to entanglement of the
photon and electron states (wave-like dressing of electrons by the
radiation) [21]. In the latter context, a promising potential application of
the RO waves for the excitation of optical nanoantennas was proposed in Ref.
[22].

Relevant to the study the propagation of the RO waves in various
nanostructures is the analysis of their nonlinearity. The modern
arsenal of nonlinear physics (harmonic generation, self-action,
multistability, solitons, \textit{etc}.), can be applied to the RO
waves in this context \cite{book}. Due to the specific physical
nature of these waves of quantum transitions, one may expect
particular manifestations of the nonlinear effects. One of the
nonlinearity mechanisms, as mentioned above, is based on the action
of local fields. This, in particular, leads to the appearance of
solitons and dynamical self-localized modes (breathers). The
construction of such discrete \textit{Rabi solitons} and
\textit{Rabi breathers} in QD arrays is the aim of the present work.
In this paper, we restrict the analysis to the semi-classical case,
i.e., the electromagnetic field is treated as a classical one, which
is quite relevant for QD settings [20]. Indeed, as, even on the
nanoscale spatial dimensions and respective femtosecond temporal
scales characteristic for these systems, quantum fluctuations of the
electromagnetic waves are negligible. The quantum description of the
photonic field may be relevant for arrays of individual atoms or
ions, rather than nanoparcticles (see, e.g., Ref. \cite{ions}).

The developed model amounts to a system of two coupled discrete nonlinear
Schr\"{o}dinger (DNLS) equations for the wave functions of the ground and
excited states, with the nonlinearity represented solely by the
cross-phase-modulation (XPM) terms, while the usual self-phase-modulation
(SPM) is absent. The system also includes the linear coupling between the
two equations, cf. Ref. \cite{coupledDNLS}. A novel essential feature of the
system, which makes it different from the previously studied ones, is the
phase shift between constants of the intersite hopping in the two components
of the system. The shift is introduced, as a geometric phase, by an angle
between the chain and the Poynting vector of the electromagnetic field,
which couples the ground and excited states of the two-level QDs. Dynamical
modes in this system are studied by means of numerical methods, which reveal
the existence of standing and moving Rabi solitons and breathers.

The paper is structured as follows. In Sec. II the model and corresponding
system of DNLS equations are formulated, and their physical meaning is
discussed in necessary detail. In Sec. III the results of the numerical
analysis for single-soliton and double-soliton complexes are presented and
conditions of their stability are reported. The paper is concluded by Sec.
IV.

\section{Model equations}

As said above, we consider a chain of identical QDs exposed to the classical
light wave with the electric field at the $p$-th site of the chain presented
as $E_{p}=\mathrm{Re}\left\{ E_{0}e^{i\left( kap-\omega t\right) }\right\} $%
, where $k$ is the axial wavenumber, and $a$ the spacing of the QD chain.
The QDs are assumed to be identical two-level non-dissipative systems with
energy distance $\hbar \omega _{0}$ between the excited and ground-state
electron orbitals, $\left\vert a_{p}\right\rangle $ and $\left\vert
b_{p}\right\rangle $, respectively. Adjacent QDs are coupled through the
electron tunneling (hopping), so that only intraband transitions are taken
into account [22]. We assume that the light interacts with the QD chain in
the resonant regime, therefore the frequency detuning is small in comparison
to both the optical and quantum-transition frequencies . Following the
rotating-wave-approximation \cite{1'}, we eliminate rapidly oscillating
terms in the equations of motion. These assumptions correspond to the RO
model formulated in Ref. \cite{20}.

The raising, lowering, and population operators of the $p$-th QD are denoted
as $\hat{\sigma}_{p}^{+}=$ $\left\vert a_{p}\right\rangle \left\langle
b_{p}\right\vert $, $\hat{\sigma}_{p}^{-}=$ $\left\vert b_{p}\right\rangle
\left\langle a_{p}\right\vert $, and $\hat{\sigma}_{zp}=$ $\left\vert
a_{p}\right\rangle \left\langle a_{p}\right\vert -\left\vert
b_{p}\right\rangle \left\langle b_{p}\right\vert $, respectively. The
corresponding single-particle Hamiltonian is
\begin{equation}
\hat{H}=\hat{H}_{0}+\hat{H}_{T}+\Delta \hat{H},  \label{HHH}
\end{equation}%
where the first term describes the QD under the action of the
electromagnetic field without tunnel coupling to its neighbors, while the
second one corresponds to the interdot coupling through the tunneling. These
two terms were included into to the model introduced in Ref. [22]. The last
term in Hamiltonian (\ref{HHH}) relates to the local-field effect. In the
framework of the Hartree-Fock-Bogoliubov approximation, it is obtained as
[10]%
\begin{equation}
\Delta \hat{H}=\frac{4\pi }{V}N_{\alpha \beta }\mu _{\alpha }\mu _{\beta
}\sum_{p}\left( \hat{\sigma}_{p}^{-}\left\langle \hat{\sigma}%
_{p}^{+}\right\rangle +\hat{\sigma}_{p}^{+}\left\langle \hat{\sigma}%
_{p}^{-}\right\rangle \right) ,  \label{1-1}
\end{equation}%
where $\mu _{\alpha }$ and $N_{\alpha \beta }$ and are, respectively,
components of the dipole-moment vector and depolarization tensor of the
single QD, $V$ is the volume of the single QD, and angle brackets denote
averaging of the corresponding operator with respect to a given quantum
state. The depolarization tensor depends both on the QD configuration and
the quantum state of the electron-hole pair:%
\begin{equation}
N_{\alpha \beta }=\frac{V}{4\pi }\int_{V}\int_{V}\left\vert \chi \left(
\mathbf{r}\right) \right\vert ^{2}\left\vert \chi \left( \mathbf{r}^{\prime
}\right) \right\vert ^{2}G_{\alpha \beta }\left( \mathbf{r}-\mathbf{r}%
^{\prime }\right) d^{3}\mathbf{r}d^{3}\mathbf{r}^{\prime },  \label{1-2}
\end{equation}%
where $\chi \left( \mathbf{r}\right) $ is the wave function of the
electron-hole, and $G_{\alpha \beta }\left( \mathbf{r}-\mathbf{r}^{\prime
}\right) $ is the Green tensor of the Maxwell's equations in the
quasi-static limit \cite{2'}.

Single-particle excitations are described by the coherent superposition,%
\begin{equation}
\left\vert \Psi (t)\right\rangle =\sum_{p}\left[ \Psi _{p}(t)e^{(i/2)\left(
kpa-\omega t\right) }\left\vert a_{p}\right\rangle +\Phi
_{p}(t)e^{-(i/2)\left( kpa-\omega t\right) }\left\vert b_{p}\right\rangle %
\right] ,  \label{1-3}
\end{equation}%
where $\Psi _{p}(t)$ \ and $\Phi _{p}(t)$ are unknown probability
amplitudes. The evolution of the system is described by the nonstationary
Schr\"{o}dinger equation for wave function $\left\vert \Psi (t)\right\rangle
$, which leads to the following system of coupled nonlinear equations for
the probability amplitudes:
\begin{eqnarray}
\frac{d\Psi _{p}}{dt} &=&iF\Psi _{p}+i\xi _{1}\left( e^{-i\phi }\Psi
_{p-1}+e^{+i\phi }\Psi _{p+1}\right)   \notag \\
&&-ig\Phi _{p}-i\Delta \omega \left\vert \Phi _{p}\right\vert ^{2}\Psi _{p},
\label{1a} \\
\frac{d\Phi _{p}}{dt} &=&-iF\Phi _{p}+i\xi _{2}\left( e^{+i\phi }\Phi
_{p-1}+e^{-i\phi }\Phi _{p+1}\right)   \notag \\
&&-ig\Psi _{p}-i\Delta \omega \left\vert \Psi _{p}\right\vert ^{2}\Phi _{p},
\label{1b}
\end{eqnarray}%
where $\phi \equiv ka/2$ is the phase shift of the oblique incident
electromagnetic wave per a half of the array period. Similar to its
counterpart in the Berry phase \cite{Berry}, this is a geometric phase, but
in the present one-dimensional setting it does not generate a topological
state. Values close to $\phi =\pi /2$ are reachable for surface waves with
large retardation \cite{21}, while $\phi =0$ corresponds to the excitation
of the array by the normally incident plane wave. Further, $g\equiv -\mu
E_{0}/\left( 2\hbar \right) $ is the QD-field coupling factor ($\mu $ is the
absolute value of the polarization vector with components $\mu _{\alpha }$),
$\Delta \omega \equiv $ $4\pi \mu _{\alpha }\mu _{\beta }N_{\alpha \beta
}/\left( \hbar V\right) $ is the depolarization shift,
\begin{equation}
F\equiv (1/2)\left( \omega -\omega _{0}\right)   \label{F}
\end{equation}%
is the an independent detuning parameter, and $\xi _{1,2}$ are coefficients
of the coupling between adjacent sites of the lattice.

These equations may be considered as the discrete version of equations that
have been used in Ref. [20] for the prediction of RO waves in a continuous
medium built of Rabi oscillators. Below, we use Eqs. (\ref{1a}) and (\ref{1b}%
) as a basic model for the analysis of RO solitons in the QD arrays. In the
simplest case, one may assume equal intersite coupling coefficients for the
ground and excited states, $\xi _{1}=\xi _{2}\equiv \xi >0$. Next, we set,
by means of an obvious rescaling, $g\equiv -1$ and $\mathrm{sign}(\Delta
\omega )\equiv -1$, hence the results are presented below in the respective
dimensionless form. In terms of general systems of coupled DNLS equations
\cite{coupledDNLS}, these signs imply, respectively, the onsite attraction
between fields $\Psi _{p}$ and $\Phi _{p}$, and the self-focusing sign of
the onsite nonlinearity. Actually, if $g$ is originally positive, it can be
made negative by substitution $\Phi _{p}\equiv -\Phi _{p}^{\prime }$, and if
$\Delta \omega $ is originally positive, it can be made negative by means of
the usual staggering substitution \cite{book}. Thus, there remain three
independent parameters in Eqs. (\ref{1a}) and (\ref{1b}), which can be
combined into frequency detuning (\ref{F}), and the complex lattice
coupling, $\xi \exp \left( i\phi \right) $, with $\phi \equiv ka/2$. We
restrict $\phi $ to interval $0\leq \phi \leq \pi /2$, while both positive
and negative values of $\omega _{0}-\omega $, i.e., of $F$, as per Eq. (\ref%
{F}) should be considered.

The cardinal difference of the present system from previously considered
models based on coupled DNLS equations \cite{book,coupledDNLS} is the
presence of the phase shift, $\phi $, in the lattice coupling. Other
differences are the mismatch parameter $F$, and the fact that the onsite
nonlinearity in Eqs. (\ref{1a}) and (\ref{1b}) is of the XPM type, with
respect to fields $\Psi _{p}$ and $\Phi _{p}$, unlike the standard DNLS
equations which feature the SPM nonlinearity \cite{book}.

Before proceeding to the detailed analysis of the model, it is relevant to
outline physical characteristics of the RO excitations described by this
setting. These are: i) the electric current and polarization induced by the
external field in the given QD; ii) the local inversion; iii) the integral
inversion. The electric polarization produced by the displacement current in
the QD via the quantum transitions is given by \cite{10}%
\begin{equation}
P_{p}=\frac{\mu }{V}\left\langle \hat{\sigma}_{p}^{-}\right\rangle +\mathrm{%
c.c.}=\frac{\mu }{V}\Psi _{p}\Phi _{p}^{\ast }e^{i\left( kpa-\omega t\right)
}+\mathrm{c.c.},  \label{P}
\end{equation}%
where $\mathrm{c.c.}$ stands for the complex conjugate. Another component of
the electrical current in the QD chain represents the electron-hole hopping
between adjacent QDs. It appears as a result of changing the probability for
the electron-hole pair to be found in the QD under the action of the
electromagnetic field. Therefore, it may be obtained, using the continuity
relation for wave function (\ref{1-3}) \cite{16}, as%
\begin{equation}
j_{p}^{T}=\frac{ie\hbar }{2ma}\left[ \Psi _{p}\left( e^{i\phi }\Psi
_{p+1}^{\ast }-e^{-i\phi }\Psi _{p-1}^{\ast }\right) +\Phi _{p}\left(
e^{-i\phi }\Phi _{p+1}^{\ast }-e^{i\phi }\Phi _{p-1}^{\ast }\right) \right] +%
\mathrm{c.c.}  \label{j}
\end{equation}%
The local inversion is defined as $\left\langle \hat{\sigma}%
_{zp}\right\rangle $, while the integral inversion, $W$, is a result of the
summation of local inversions over the chain:%
\begin{equation}
W=\sum_{p}\left( \left\vert \Psi _{p}\right\vert ^{2}-\left\vert \Phi
_{p}\right\vert ^{2}\right) .  \label{W}
\end{equation}%
The integral inversion shows the difference for the QD chain to be found in
the excited and ground states, respectively. This value varies in interval $%
-1\leq W\leq +1$, generalizing the correspondent characteristic of the
single atom \cite{1'}.

\begin{figure}[th]
\center\includegraphics[width=12cm]{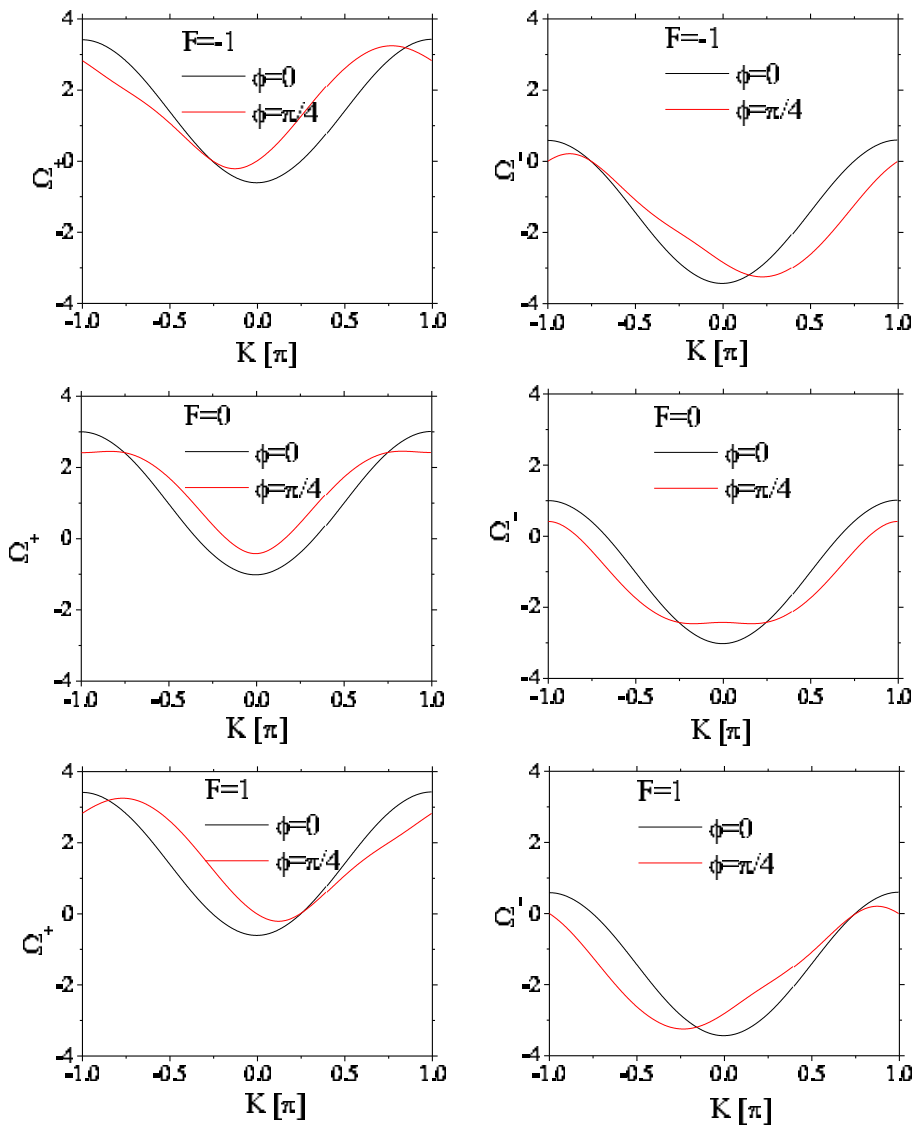}
\caption{(Color online) Dispersion curves $\Omega (K)$ for the linearized
system at different combinations of the values of parameters $F$ and $%
\protect\phi $, as indicated in the corresponding plots. Other coefficients
are $\protect\xi =1$ and $\,g=-1$. The complexes formed of fundamental
discrete solitons are expected to exist in the regions with $\Omega <0$
(semi-infinite gaps), i.e., below the black (red) lines, depending on the
system's parameters.}
\label{dispersion}
\end{figure}

The dispersion relation for Eqs. (\ref{1a}), (\ref{1b}) can be derived by
looking for solutions to the linearized version of the system as
\begin{equation}
\left\{ \Psi _{p},\Phi _{p}\right\} =\left\{ A,B\right\} \exp \left(
iKp-i\Omega t\right) .
\end{equation}%
A straightforward analysis yields two branches of the dispersion relation,
\begin{equation}
\Omega =-2\xi \left( \cos \phi \right) \cos K\pm \sqrt{g^{2}+\left[ F-2\xi
\left( \sin \phi \right) \sin K\right] ^{2}}  \label{disp}
\end{equation}%
[recall that we actually fix $g\equiv -1$, and $F$ is defined as per Eq. (%
\ref{F})], examples of which are shown in Fig. \ref{dispersion}. In the
limit of $\phi =0$, Eq. (\ref{disp}) goes over into the known relation for
the usual system of linearly coupled DNLS equations, cf. Ref. \cite%
{coupledDNLS}, which consists of two similar branches shifted by a constant,
$\Delta \Omega =2\sqrt{g^{2}+F^{2}}$. In the case of $\phi \neq 0$, $\Delta
\Omega $ is no longer a constant, depending on wavenumber $K$.

The objective of the work is to construct discrete solitons in this model,
and then test their dynamical stability and mobility. Obviously, the
solitons must be located inside gaps of the dispersion relation (\ref{disp}%
), see Fig. \ref{dispersion}.

Stationary soliton solutions of Eqs. (\ref{1a}) and (\ref{1b}) were obtained
by adopting the nonlinear equation solver based on the Powell method \cite%
{powell}. Direct dynamical simulations were based on the Runge-Kutta
numerical procedure of the sixth order \cite{powell}. The numerical solution
was performed for a finite lattice, $-N/2\leq p\leq +N/2$, of size $N+1$
with even $N$, and periodic boundary conditions. In this case, the system
preserves the total probability, which is normalized to be $1$:
\begin{equation}
P=\sum_{p=-N/2}^{+N/2}(|\Psi _{p}|^{2}+|\Phi _{p}|^{2})\equiv 1,
\label{power}
\end{equation}%
and the corresponding energy,%
\begin{gather}
\left\langle \hat{H}\right\rangle =\frac{1}{2}\sum_{p=-N/2}^{+N/2}\left\{
-F\left( \left\vert \Psi _{p}\right\vert ^{2}-\left\vert \Phi
_{p}\right\vert ^{2}\right) \right.  \notag \\
-\xi \Psi _{p}^{\ast }\left( \Psi _{p-1}e^{-i\phi }+\Psi _{p+1}e^{i\phi
}\right) -\xi \Phi _{p}^{\ast }\left( \Phi _{p-1}e^{i\phi }+\Phi
_{p+1}e^{-i\phi }\right)  \notag \\
\left. +g\Psi _{p}^{\ast }\Phi _{p}+\frac{1}{2}\Delta \omega \left\vert \Psi
_{p}\right\vert ^{2}\left\vert \Phi _{p}\right\vert ^{2}\right\} +\mathrm{%
c.c.}  \label{H}
\end{gather}%
The analysis presented below was performed with $N=101$, unless stated
otherwise.

Before proceeding to the description of results, it is relevant to discuss
the applicability of the proposed model to really existing QDs. Parameters
of an isolated QD, as well as the inter-dot coupling, widely vary depending
on the configuration, material and environment of the QD. Coefficients which
are used below are realistic for typical semiconductor-based QDs. Firstly,
as noted in Ref. \cite{15}, for the observation of the RO modified by the
local field effects, it is essential to use an exciting optical pulse with $%
\Omega \lesssim \Delta \omega $. This implies that one should use an optical
pulse which is longer than the Rabi-period, $2\pi /\Omega $, but still
sufficiently shorter than transverse damping time, $T_{\mathrm{damp}}$. Such
conditions where experimentally realized, in particular, in Ref. \cite{15}
(with $T_{\mathrm{damp}}\simeq 2.5$ ns) for the QDs grown in a capped layer.
Further, the general condition, $g\sim \xi $, defines a range of the
field-QD coupling strength in which the interplay of the Rabi dynamics and
interdot tunneling is essential. As shown in Ref. \cite{21}, this condition
is realistic in typical experimental situations.

\section{Results and discussion}

\subsection{Single-soliton complexes}

Stationary solitons with frequency $\Omega $ are looked for in the usual
form,%
\begin{equation}
\left\{ \Psi _{p}(t),\Phi _{p}(t)\right\} =e^{-i\Omega t}\left\{
A_{p},B_{p}\right\} .  \label{AB}
\end{equation}%
Two-component single-soliton complexes are composed of single-peak pulses in
fields $A_{p}$ and $B_{p}$. Recall that, in the single-component DNLS
equation, there are two types of fundamental discrete solitons, stable
onsite-centered and unstable intersite-centered ones \cite{book}. We have
found that Eqs. (\ref{1a}) and (\ref{1b}) give rise to two species of
stationary single-soliton complexes, in which both components are of the
same type, either onsite or intersite. Accordingly, these species are
referred to as SOOM (single on-on-site mode) and SIIM (single
inter-inter-site mode) solutions, respectively. As concerns the relative
sign of the two components, the complexes characterized by
\begin{equation}
\mathrm{sgn}\left\{ \mathrm{Re}(A_{p})\mathrm{Re}(B_{p})\right\} =\left\{
+1,-1\right\}   \label{+-}
\end{equation}%
are called in-phase and counter-phase configurations, respectively.

Stable stationary solitons and robust breathers found in present
model are reported below. Before proceeding to their detailed
description, it is relevant to discuss the physical purport of
these localized collective excitations. Firstly, we note that Eqs.
(\ref{1a}) and (\ref{1b}) are invariant with respect to
substitution $i\rightarrow -i,~\Psi _{p}\longleftrightarrow \Phi
_{p}$. This implies the existence of a set of SIIM and SOOM
solutions satisfying conditions $\left\vert A_{p}\right\vert
=\left\vert B_{p}\right\vert $, $\mathrm{Re}\left( A_{p}\right) =\mathrm{Re}%
\left( B_{p}\right) $, $\mathrm{Im}\left( A_{p}\right) =-\mathrm{Im}\left(
B_{p}\right) $. Examples of dynamically stable SOOM solitons, which may be
considered as \textit{Rabi solitons}, are presented at Fig. \ref{slika1}.
They are characterized by the conserved energy, see Eq. (\ref{H}), which
includes the contribution from the QD chain proper, and the energy of
field-QD interaction. Actually, quasiparticles represented by the Rabi
solitons may be considered as artificial RO atoms. Dynamical polarization (%
\ref{P}) of such excitations harmonically oscillates at the frequency of the
external field. These oscillations do not manifest themselves in observable
values (such as induced currents), rather serving for the inner
stabilization of the Rabi soliton. The tunneling transitions are able to
produce the dc current in the soliton, as per Eq. (\ref{j}). Its value is
proportional to $\sin \phi $, hence it vanishes at $\phi \rightarrow 0$.
Further, according to Eq. (\ref{W}), the inversion vanishes for the
stationary Rabi solitons.

The situation is different for the breathing SIIM structures, for which the
corresponding spectra feature two lines with frequencies $\nu =\Omega $ and $%
\nu \approx \Omega /2$, see Fig. \ref{slika2}(d) below. This entails
emergence of additional lines in the current spectra. In particular, the
resonant line at the external-field frequency transforms into a triplet with
frequencies $\nu =\Omega $, $\nu \approx \Omega \pm \Omega /2$. Its physical
nature is similar to the well-known Mollow triplet in the resonant
fluorescence \cite{1'}. The dc current transforms into a low-frequency part
of the current spectrum, which partially encloses the resonance line at
frequency $\nu =\Omega /2$. The appearance of this spectral feature is
stipulated by the asymmetry produced by the phase shift. The physical
mechanism of its creation is similar to that for the spectral line at the
Rabi frequency in the QD's RO with spatially broken symmetry, predicted in
Ref. \cite{1''}.

\subsubsection{The case of $\protect\phi =0$}

\begin{figure}[th]
\center\includegraphics[width=14cm]{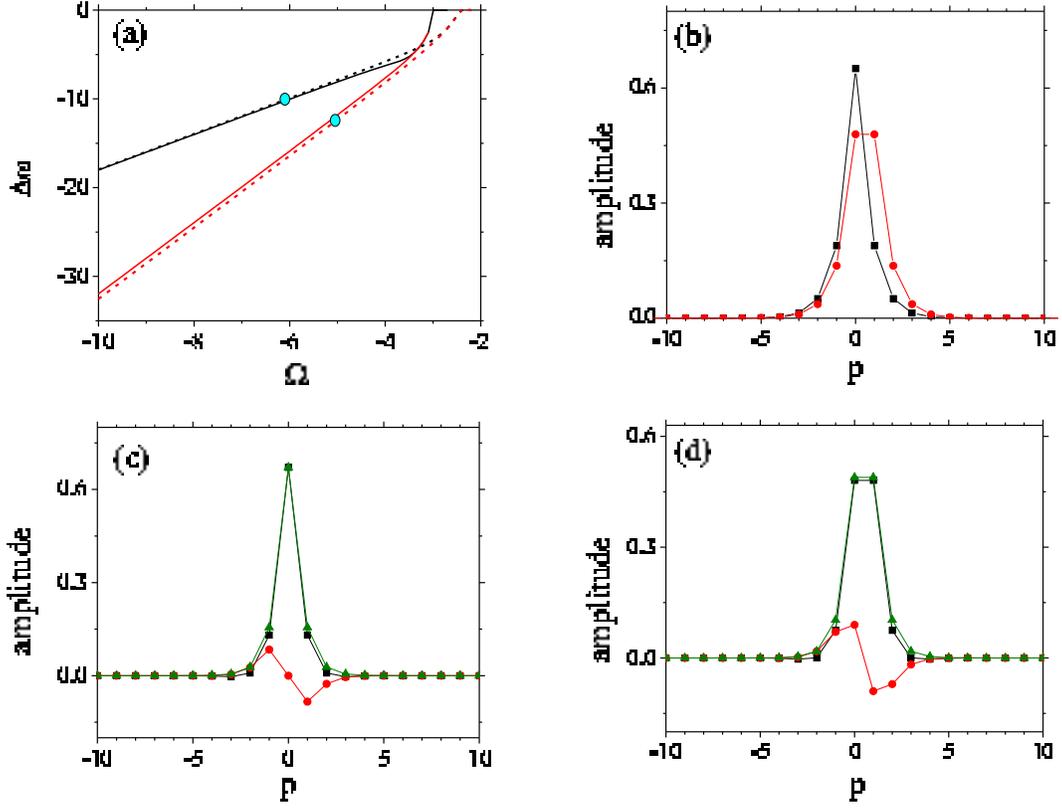}
\caption{(Color online) (a) The $\Delta \protect\omega $ vs. $\Omega $
diagrams for SOOM and SIIM (black and red line, respectively). Solid lines
correspond to $\protect\phi =0$, while dashed ones to $\protect\phi =\protect%
\pi /4$. (b) Amplitude profiles of onsite components of the SOOM (black) and
intersite components of SIIM (red) ($\protect\phi =0$). Plots (c) and (d)
show real and imaginary parts (black and red lines) and the absolute value
(green) of one component of SOOM and SIIM, respectively, for $\protect\phi =%
\protect\pi /4$. Note relations $|A_{p}|=|B_{p}|$, \thinspace $\mathrm{Re}%
(A_{p})=\mathrm{Re}(B_{p})$, and $\mathrm{Im}(A_{p})=-\mathrm{Im}(B_{p})$
between the components of these solitons.}
\label{slika1}
\end{figure}

In the case of $\phi =0\ $and $F=0$ [no frequency mismatch, see Eq. (\ref{F}%
)], stationary single-soliton complexes are formed of two identical onsite
or intersite solitons, which are standard DNLS modes with real probability
amplitudes, $\mathrm{Im}(A_{p})=\mathrm{Im}(B_{p})=0$, see Fig. \ref{slika1}%
(b). Two configurations, in-phase and counter-phase, of each complex differ
by the sign of the relative sign of the two components. Families of such
complexes are represented by the corresponding $\Delta \omega (\Omega )$
curves in Fig. \ref{slika1}(a). It is relevant to stress that families of
the solitons in the present system are characterized by such curves (the
nonlinearity coefficient versus the carrier frequency), as the total
probability (soliton's norm) is fixed as per Eq. (\ref{power}). The
situation is opposite in nonlinear optics, where the nonlinearity
coefficient is fixed, while the total norm varies, showing the total power
of the soliton \cite{book}. Of course, as concerns actual solutions of the
model, these two settings may be transformed into each other by rescaling.

\begin{figure}[th]
\center\includegraphics [width=12cm]{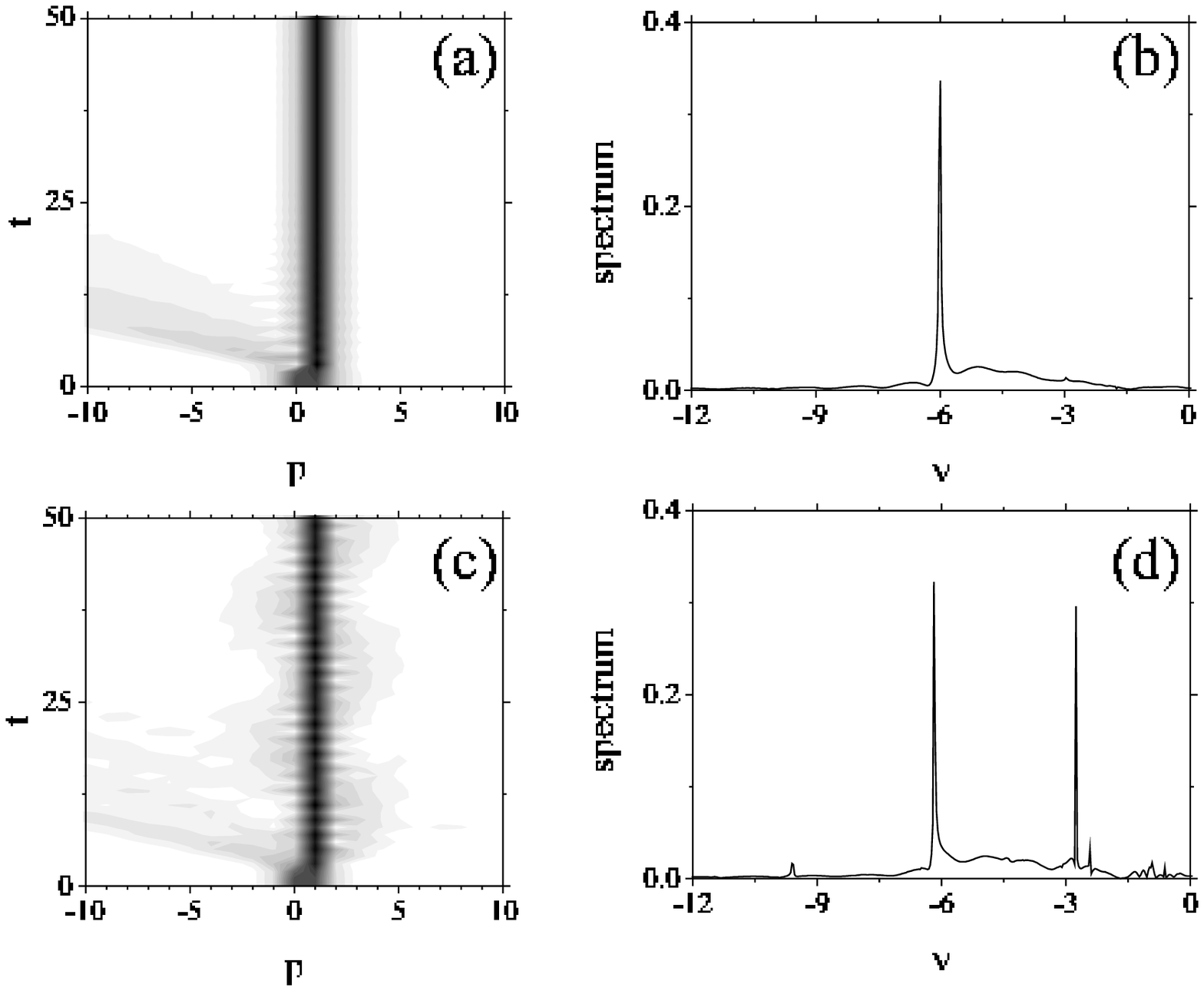}
\caption{Examples of unstable evolution of SIIM. One of the components is
displayed here. (a) The SIIM with $\Omega =-5,\Delta \protect\omega =-11.81$
and $\protect\phi =0$ transforms into an SOOM breather with smaller $\Delta
\protect\omega =-10.3$, and $\Omega =-6$ (circles on the curves). The
corresponding breather's spectrum is shown in (b). (c) SIIM with $\Omega
=-5,\Delta \protect\omega =-12.32,\protect\phi =\protect\pi /4$ transforms
into an onsite breather, whose spectrum is shown in (d).}
\label{slika2}
\end{figure}

In accordance with the above-mentioned stability properties of the usual
DNLS solitons, only the SOOM of the in-phase type is found to be dynamically
stable, while the SIIM\ evolves, due the instability, into a localized
breathing structure. An example of the evolution of an unstable SIIM, with $%
\Delta \omega =-11.81,\Omega =-5$, into a periodic breather of the SOOM type
(with $\Omega \approx -6$) is displayed in Fig. \ref{slika2}. The
corresponding spectrum in Fig. \ref{slika2}(c) features one characteristic
peak, at $\nu \approx -6$.

Similar behavior is found in the parameter region where the SIIM and SOOM
complexes with close values of $\Delta \omega $ coexist [an area of small $%
|\Omega |$ in Fig. \ref{slika1}(a)]: an unstable SIIM radiates away a small
part of its norm and transforms themselves into a SOOM [circles on the
respective curves in Fig. \ref{slika1}(a) designate the initial and final
states corresponding to the example displayed in Fig. \ref{slika2}(a)]. The
spontaneous transformation into a quasi-periodic breather is found too in
the region with well separated $\Delta \omega (\Omega )$ curves for
different species of the single-soliton complexes. The breathers feature a
more complex structure if they develop from the counter-phase original
configurations.

\subsubsection{The case of $\protect\phi \neq 0$}

\begin{figure}[th]
\center\includegraphics [width=12cm]{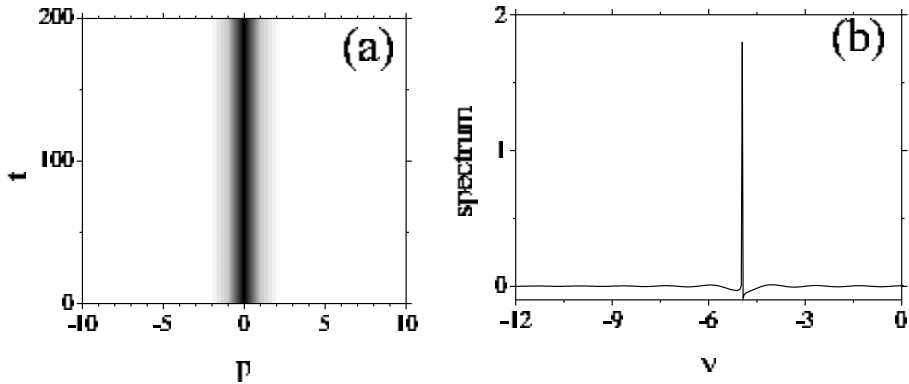}
\caption{(a) An example of the evolution of the dynamically stable SOOM with
$\Omega =-5$, $\Delta \protect\omega =-7.88$ and $\protect\phi =\protect\pi %
/4$. (b) The corresponding spectrum is characterized by the single line at $%
\protect\nu =\Omega $. }
\label{slika3}
\end{figure}

As said above, the central issue addressed in this work is the novel type of
the system of coupled DNLS equations (\ref{1a}) and (\ref{1b}) with $\phi
\neq 0$. In this case, the onsite and intersite modes are characterized by
complex amplitudes, with $\mathrm{Im}(A_{p}),\mathrm{Im}(B_{p})\neq 0$, see
Figs. \ref{slika1}(c) and (d). As in the previous case, the in-phase and
counter-phase configurations of the single-soliton modes can be
distinguished, as per definition (\ref{+-}). In both configurations, the
relative sign of the imaginary parts of the amplitudes is opposite to that
for the real parts. Similar to the case of $\phi =0$, only the in-phase
SOOMs are stable in the present case, see Fig. \ref{slika3}, while SIIMs
develop into breathers of the onsite type, as shown in Figs. \ref{slika2}(c)
and (d).

The presence of the nonzero frequency detuning, $F\neq 0$ [see Eq. (\ref{F}%
)] makes the two components of SOOM and SIIM solutions mutually asymmetric.
All the findings concerning the dynamical stability of the single-soliton
complexes remain the same as for $F=0$. This conclusion turns out to be
universal in the framework of the present model, therefore all the results
are presented for $F=0$.

\begin{figure}[th]
\center\includegraphics [width=8cm]{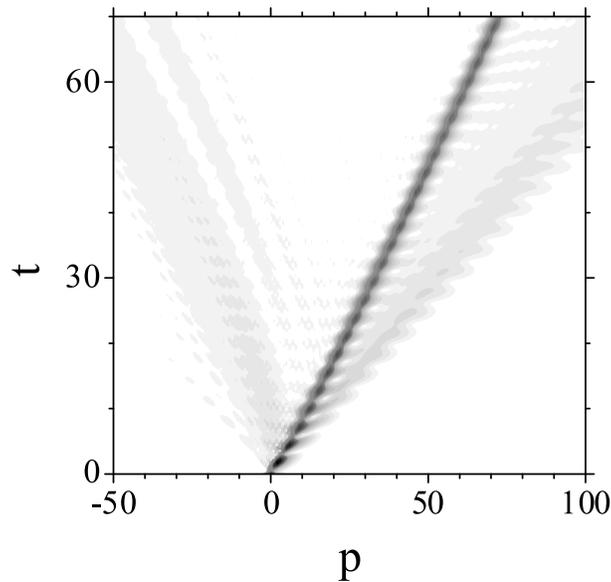}
\caption{An example of the moving breathing component of the dynamically
stable SOOM with $\Omega =-3.6$, $\Delta \protect\omega =-5.86472,\,N=301$
and $\protect\phi =0$. The kick strength is $K=\protect\pi /6$.}
\label{mov}
\end{figure}

Concluding this subsection, we note that systematic simulations demonstrate
that moving single-soliton complexes are found in a robust form only in the
region where SOOM and SIIM species coexist with nearly equal values of $%
\Delta \omega $. As usual, the moving modes can be generated, for
any value of $\phi $, from stable standing ones by the application
of the kick to both components, $\left\{ \Psi _{p},\Phi
_{p}\right\} \rightarrow e^{iKp}\left\{ \Psi _{p},\Phi
_{p}\right\} $. However, it is observed that with increasing the
values of $\phi$ the moving mode can be trapped by the lattice
potential. Physically, this means that the soliton is created in a
segment of the QD chain where an additional ramp (linear
potential) is applied. Alternatively, the kick may be imparted by
an additional nonresonant laser beam shone along the chain. The
resulting running solitons seem as breathing complexes traveling
across the lattice, see Fig. (\ref{mov}).

\subsection{Double soliton complexes}

\subsubsection{Building bound complexes}

\begin{figure}[th]
\center\includegraphics [width=12cm]{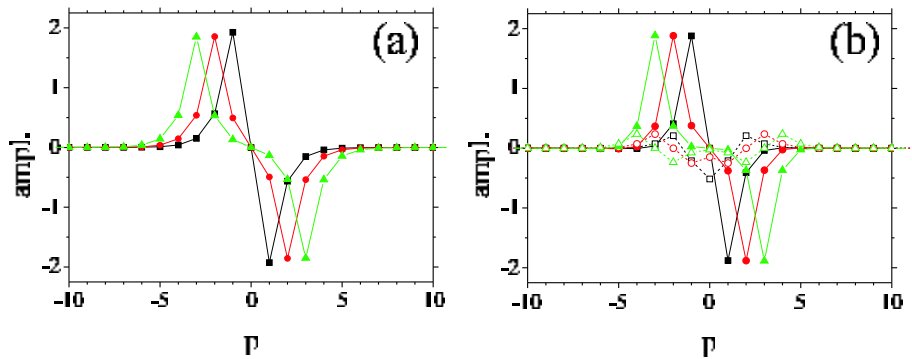}
\caption{(Color online) (a) The antisymmetric bound state of two onsite
solitons (DOOM), with separations $\Delta =1,3,5$ between centers of the
solitons, and $\protect\phi =0,\Omega =-5$. (b) The real part (full symbols
and solid lines) and imaginary part (empty symbols and dashed lines) for
component of the antisymmetric DOOM with $\Omega =-5$ and $\protect\phi =%
\protect\pi /4$. Black, red and green lines correspond to $\Delta =1,3,5$,
respectively}
\label{slika4}
\end{figure}

\begin{figure}[th]
\center\includegraphics [width=12cm]{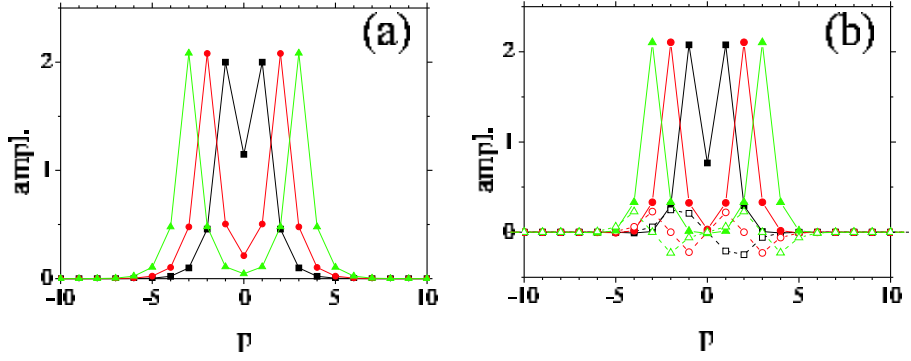}
\caption{(Color online) (a) The same as in Fig. \protect\ref{slika4}(a) but
for a symmetric bound state (DOOM), with $\Omega =-5.8$. (b) The real part
(full symbols and solid lines) and imaginary part (empty symbols and dashed
lines) for the component of the symmetric DOOM with $\Omega =-5.8$ and $%
\protect\phi =\protect\pi /4$. Black, red and green lines correspond to $%
\Delta =1,3,5$. }
\label{slika6}
\end{figure}

The same system of Eqs. (\ref{1a}) and (\ref{1b}) makes it possible to
create two-component double-soliton complexes (bound states of fundamental
discrete solitons), with two peaks in each component. Stable among them may
naturally be double on-on-site modes (DOOMs, i.e., bound states of SOOMs).
Here we analyze the dynamics of DOOMs built of two fundamental soliton
complexes with the separation between their centers equal to one, three, and
five lattice periods ($\Delta =1,3,5$), see Figs. \ref{slika4} and \ref%
{slika6}. Depending on the symmetry with respect to the central lattice
site, antisymmetric (odd) and symmetric (even) DOOMs can be identified, see
Figs. \ref{slika4} and \ref{slika6}, respectively. In particular, the
antisymmetric bound state with $\Delta =1$ corresponds, in the case of the
usual single-component DNLS equation, to the well-known twisted modes \cite%
{twisted}. Here, we consider only DOOMS of the in-phase type, as concerns
the relative phase of the two components, because counter-phase bound states
turnout to be strongly unstable.

\begin{figure}[th]
\center\includegraphics [width=12cm]{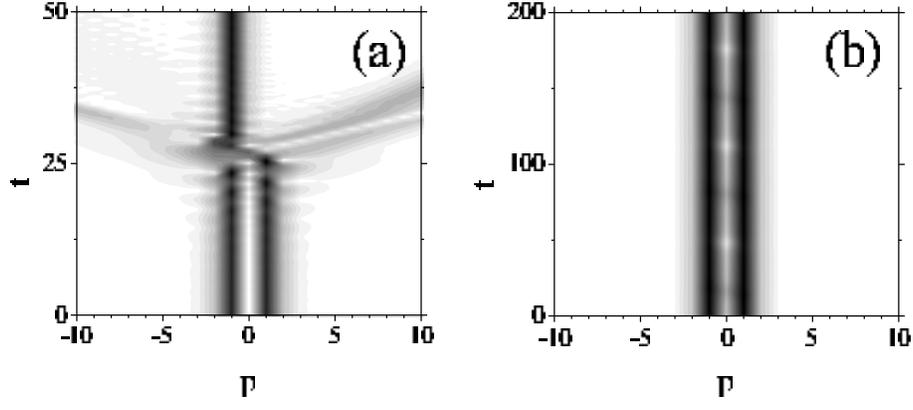}
\caption{A breathing localized configuration which is a part of the complex
formed by perturbing the antisymmetric DOOM with $\Delta =1,\protect\phi %
=0,\Omega =-5,\Delta \protect\omega =-16.2$ (a), and $\Delta =1,\protect\phi %
=\protect\pi /4,\Omega =-5,\Delta \protect\omega =-15.7$ (b). }
\label{slika5}
\end{figure}

\begin{figure}[th]
\center\includegraphics [width=21cm]{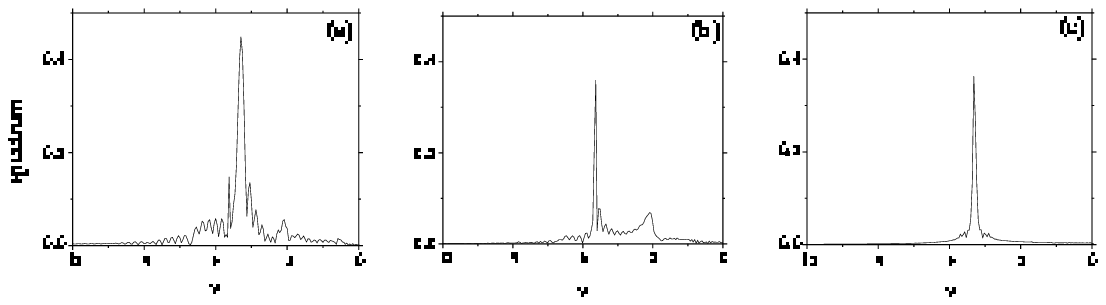}
\caption{Spectra of the breathers shown in Fig. \protect\ref{slika5}. Plots
(a) and (b) show the power spectra at sites $p=-1$ and $p=1$ for $\Delta =1,%
\protect\phi =0,\Omega =-5,\Delta \protect\omega =-16.2$, which corresponds
to Fig. \protect\ref{slika5}(b). Plots (c) shows the spectrum at $p=-1$ for $%
\Delta =1,\protect\phi =\protect\pi /4,\Omega =-5,\Delta \protect\omega %
=-15.7$, which corresponds to Fig. \protect\ref{slika5}(b). Both
components of the double soliton complexes are identical, therefore
the spectra are shown for one of them.} \label{slika8}
\end{figure}

\begin{figure}[th]
\center\includegraphics [width=12cm]{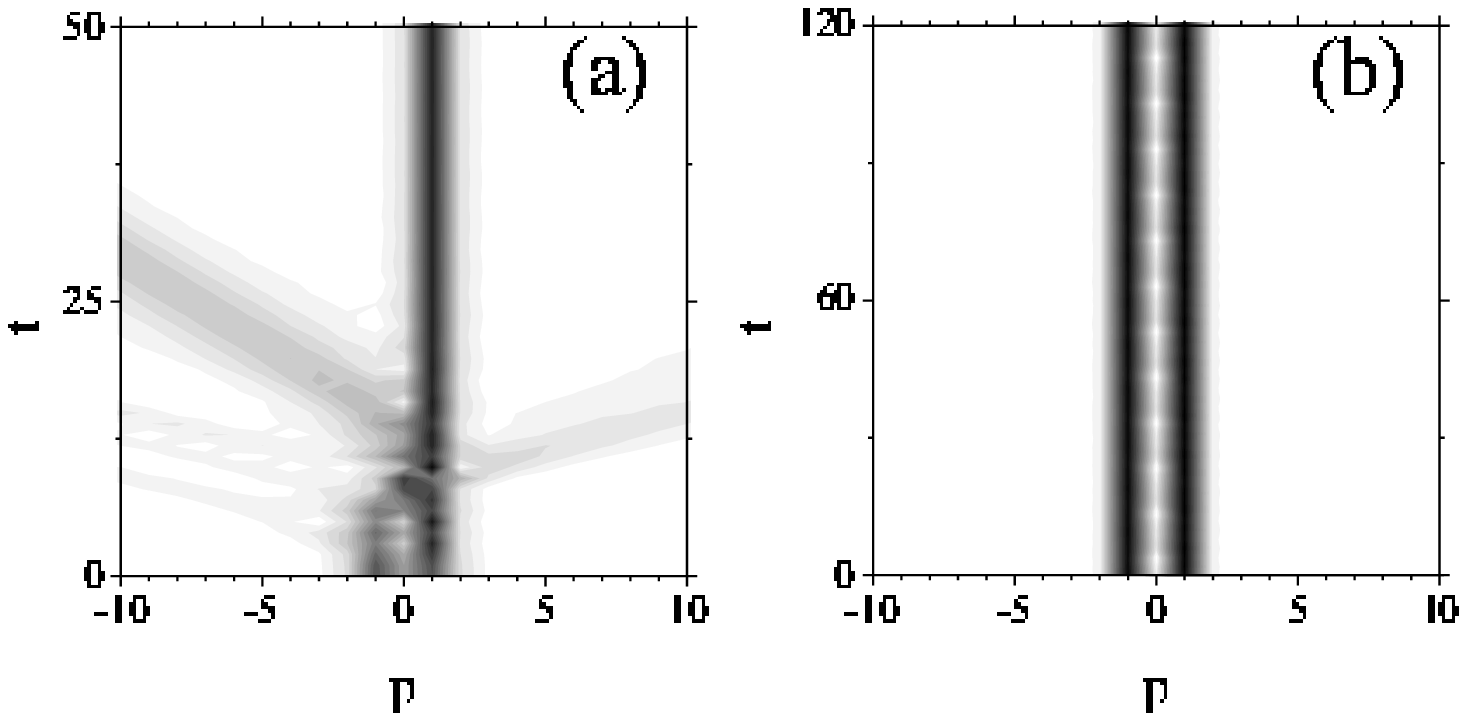}
\caption{The breather which is part of the complex formed by a perturbed
symmetric DOOM with $\Delta =1,\protect\phi =0,\Omega =-5.8$ (a), and $%
\Delta =1,\protect\phi =\protect\pi /4,\Omega =-5.8$ (b). }
\label{slika7}
\end{figure}

\subsubsection{The case of $\protect\phi =0$}

The amplitudes of the components of the stationary DOOMs in this case are
real, as it was the case for SOOMs. Apart from the symmetric DOOM with $%
\Delta =1$, which is strongly unstable in its whole existence region,
complexes of other types are stable, except for narrow instability regions,
which shrinks with the increase of separation $\Delta $ between the two
solitons. The situation is similar to its well-known counterpart for the
single DNLS equation \cite{Panos}. Examples of the instability development
are shown in Figs. \ref{slika5}(a) and \ref{slika7}(a) for antisymmetric and
symmetric DOOMs, respectively. To illustrate the structure of the breather
which is presented in Fig. \ref{slika5} (a), its power spectra at sites $p=-1
$ and $p=1$ are shown in Fig. \ref{slika8}.

\subsubsection{The case of $\protect\phi \neq 0$}

In the system with $\phi \neq 0$, both antisymmetric and symmetric DOOMs
evolve into bound states of onsite breathers featuring periodic or
quasiperiodic oscillations, depending on separation $\Delta $ and parameters
of the system, see Figs. \ref{slika5}(b) and \ref{slika7}(b). The spectrum
of the on-site breather at site $p=-1$, for the case presented in Fig. \ref%
{slika5}(b), features a single peak centered close to $\nu =-5$ (not shown
here). For other $p$, the corresponding spectra are centered at the same
frequency.

\section{Conclusions}

Our objective in this work was to develop the theory for the interaction of
the QD (quantum--dot) chain with classical light in the strong-coupling
regime, which takes into account the local-field effect (depolarization).
The latter effect makes the system nonlinear. We have introduced the
periodic array of identical two-level QDs, coupled via tunneling between
neighboring dots. The system of equations of motion has been derived, which
includes the tunneling and depolarization. The main conclusions produced by
the analysis can be summarized as follows.

(i) The equations of motion have the form of two coupled DNLS\
(discrete nonlinear Schr\"{o}dinger) equations for probability
amplitudes of the given QD (lattice site) to be found in the ground
or excited state. There are two different mechanisms of coupling
between the equations. The linear mechanism is stipulated by the
``chain-light" interaction, and manifests itself in the form of the
RO (Rabi oscillations). The other coupling mechanism is based on the
XPM nonlinear terms, which represent the action of the local fields.
The constants of the intersite lattice coupling (hopping) are
complex ones, whose phases, which are opposite in the two coupled
DNLS equations, are determined by the angle between the chain's axis
and the direction of the incident electromagnetic wave. These
intersite hopping coefficients, which are complex conjugate to each
other in the two equations, represent a drastic difference of the
present model from previously studied systems of coupled DNLS
equations.

(ii) The numerical analysis has demonstrated that RO waves in the QD chain,
coupled to the co-propagating classical electromagnetic field, self-trap
into stable onsite-centered fundamental \textit{Rabi solitons}, while the
intersite fundamental solitons are unstable, spontaneously transforming into
onsite-centered robust \textit{Rabi breathers}. Bound states of fundamental
solitons are unstable in the stationary state, but they too readily
transform themselves into robust breathers. The above-mentioned intersite
hopping coefficients, which take complex conjugate in the coupled equations,
affect the internal structure of the solitons, making them complex modes,
but this novel feature does not destabilize the solitons. The detuning
between the light and inter-level transition frequencies breaks the symmetry
between the two components of the discrete solitons, but does not affect
their stability either. Mobility of the Rabi solitons was investigated too.

(iii) The frequency spectrum of the stable Rabi solitons features a single
narrow line at the Rabi frequency. The RO at this frequency do not manifest
themselves in terms of observable quantities, serving for the stabilization
of the solitons. The inversion of the stable Rabi soliton [$W$, see Eq. (\ref%
{W})] is exactly equal to zero. The frequency spectrum of electric current
in the QD chain comprises a narrow line at the external field frequency, as
well as the zero-frequency line corresponding to the dc current. The former
frequency is produced by the displacement current due to the resonant
quantum transitions, while the latter one is induced by the tunneling
(intersite hopping).

(iv) The frequency spectrum of the robust Rabi breathers features an
additional spectral line, which produces extra lines in the current spectra.
The resonant line at the external-field frequency transforms into a triplet,
while the dc current transforms into the low-frequency spectral component,
which partially overlaps with the single resonance line. The low-frequency
current component is induced by the asymmetry, which is determined by the
direction of the light propagation relative to the QD chain. Accordingly,
the low-frequency component vanishes in the limit of the normal incidence.

The theory presented in this paper does not account for relaxation
mechanisms, the quantum structure of the incident light, and interaction of
moving Rabi-solitons with edges of the chain (unless the chain is circular,
and has no edges). These problems are of special interest and should be
considered elsewhere.

\acknowledgments G.G., A.M., and Lj.H. acknowledge support from the Ministry
of Education and Science of Serbia (Project III45010).

\end{document}